\documentclass[journal]{IEEEtran}
 \pdfoutput=1
 \usepackage{cite}
\usepackage{amsmath,amssymb,amsfonts}
\usepackage{algorithmic}
\usepackage{graphicx}
\usepackage{textcomp}
\usepackage{framed,multirow}
\usepackage{latexsym}
\usepackage{amsmath}
\usepackage{threeparttable}
\usepackage{float}
\usepackage{array}
\usepackage{hhline}
\usepackage{colortbl}
\usepackage{stfloats}
\usepackage{booktabs}
\usepackage{multirow}
\usepackage{enumerate}
\usepackage{diagbox}
\usepackage{url}
\usepackage{hyperref}
\usepackage{makecell,rotating}

\def\etal{\textit{et~al}.}

\DeclareGraphicsExtensions{.pdf,.jpeg,.png,.jpg}

\graphicspath{{fig/}}

\definecolor{mygray}{gray}{.9}

\definecolor{newcolor}{rgb}{.8,.349,.1}

\makeatletter

\newcommand{\Rmnum}[1]{\expandafter\@slowromancap\romannumeral #1@}
\makeatother

\title{HoVer-Trans: Anatomy-aware HoVer-Transformer for ROI-free Breast Cancer Diagnosis in Ultrasound Images}
\author{Yuhao Mo, Chu Han, \IEEEmembership{Member, IEEE}, Yu Liu, Min Liu, Zhenwei Shi, Jiatai Lin, Bingchao Zhao, Chunwang Huang, Bingjiang Qiu, Yanfen Cui, Lei Wu, Xipeng Pan, Zeyan Xu, Xiaomei Huang, Zaiyi Liu, Ying Wang, Changhong Liang
\thanks{Yuhao Mo and Bingchao Zhao are with the School of Medicine, South China University of Technology, Guangzhou 510006, Guangdong, China; the Department of Radiology, Guangdong Provincial People’s Hospital, Guangdong Academy of Medical Sciences, Guangzhou, 510080, China;}
\thanks{Changhong Liang is with the Department of Radiology, Guangdong Provincial People’s Hospital, Guangdong Academy of Medical Sciences, Guangzhou, 510080, China; the School of Medicine, South China University of Technology, Guangzhou 510006, Guangdong, China.}
\thanks{Chu Han, Yu Liu, Zhenwei Shi, Bingjiang Qiu, Lei Wu, Xipeng Pan, Zeyan Xu, Xiaomei Huang and Zaiyi Liu are with the Department of Radiology, Guangdong Provincial People’s Hospital, Guangdong Academy of Medical Sciences, Guangzhou, 510080, China;
Guangdong Provincial Key Laboratory of Artificial Intelligence in Medical Image Analysis and Application, Guangdong Provincial People's Hospital, Guangdong Academy of Medical Sciences, Guangzhou, 510080, China.}
\thanks{Jiatai Lin is with the School of Computer Science and Engineering, South China University of Technology, Guangzhou, 510006, China.}
\thanks{Chunwang Huang is with the Department of Ultrasound, Guangdong Provincial People’s Hospital, Guangdong Academy of Medical Sciences, Guangzhou, 510080, China}
\thanks{Ying Wang is with the Department of Medical Ultrasonics, The First Affiliated Hospital of Guangzhou Medical University, Guangzhou, Guangdong, China.}
\thanks{Min Liu is with the Department of Ultrasound, Sun Yat-sen University Cancer Center; State Key Laboratory of Oncology in South China; Collaborative Innovation Center for Cancer Medicine, Guangzhou 510060, China.}
\thanks{Corresponding author: Zaiyi Liu, Ying Wang, Changhong Liang.}
\thanks{The first four authors contributed equally.}}

\begin{document}
\maketitle

\IEEEtitleabstractindextext{\begin{abstract}
Ultrasonography is an important routine examination for breast cancer diagnosis, due to its non-invasive, radiation-free and low-cost properties. However, the diagnostic accuracy of breast cancer is still limited due to its inherent limitations. It would be a tremendous success if we can precisely diagnose breast cancer by breast ultrasound images (BUS). Many learning-based computer-aided diagnostic methods have been proposed to achieve breast cancer diagnosis/lesion classification. However, most of them require a pre-define ROI and then classify the lesion inside the ROI. Conventional classification backbones, such as VGG16 and ResNet50, can achieve promising classification results with no ROI requirement. But these models lack interpretability, thus restricting their use in clinical practice. In this study, we propose a novel ROI-free model for breast cancer diagnosis in ultrasound images with interpretable feature representations. We leverage the anatomical prior knowledge that malignant and benign tumors have different spatial relationships between different tissue layers, and propose a HoVer-Transformer to formulate this prior knowledge. The proposed HoVer-Trans block extracts the inter- and intra-layer spatial information \textit{horizontally} and \textit{vertically}. We conduct and release an open dataset \textit{GDPH\&SYSUCC} for breast cancer diagnosis in BUS. The proposed model is evaluated in three datasets by comparing with four CNN-based models and two vision transformer models via five-fold cross validation. It achieves state-of-the-art classification performance with the best model interpretability. In the meanwhile, our proposed model outperforms two senior sonographers on the breast cancer diagnosis when only one BUS image is given.
\end{abstract}
\begin{IEEEkeywords}
Breast cancer diagnosis, Transformer, Ultrasound, Anatomical structure
\end{IEEEkeywords}}

\maketitle
\IEEEdisplaynontitleabstractindextext

\IEEEpeerreviewmaketitle

\vspace{-2mm}
\section{Introduction}
Breast cancer is the most commonly diagnosed cancer and the leading cause of cancer death in women globally~\cite{sung2021global}. Early breast cancer diagnosis can reduce mortality and increase survival rates~\cite{massat2016impact}. 
Breast ultrasound (BUS) is an important imaging modality for breast cancer diagnosis and screening because it is low-cost, non-invasive, radiation-free, and relatively more sensitive for dense breast tissue~\cite{bevers2018breast}. In addition, ultrasound is effective at the differentiation of cysts from solid lesions~\cite{qian2021prospective}. Therefore, it is meaningful for breast cancer patients if there exists a precise diagnostic method for BUS, especially for the dense breast patients in Asia.

Currently, BUS evaluation generally relies on the subjective evaluation of sonographers. However, the diagnostic accuracy of ultrasound is constrained by the limited number of specialized sonographers. In addition, a high intra- and inter-observer variability exists even among expert sonographers. To overcome such difficulties, computer-aided diagnosis (CAD) systems~\cite{shen2021artificial,wang20203d,wang2019deeply} have been constructed to help sonographers with a more efficient and more precise breast cancer diagnosis. With the recent advancements of deep learning models, the performance of the diagnostic models even outperforms expert sonographers~\cite{qian2020combined}. Even though existing models have already achieved outstanding diagnostic performance, most of them are still a `black box', which lacks interpretability. Furthermore, since the open-source data in the community is very limited, existing models either evaluated the performance on a relatively small open dataset (BUSI~\cite{al2020dataset} or UDIAT~\cite{yap2017automated}) or their private datasets. The clinical usability of the diagnostic model should be further assessed.

\begin{figure}
	\centering
	\includegraphics[width=.99\linewidth]{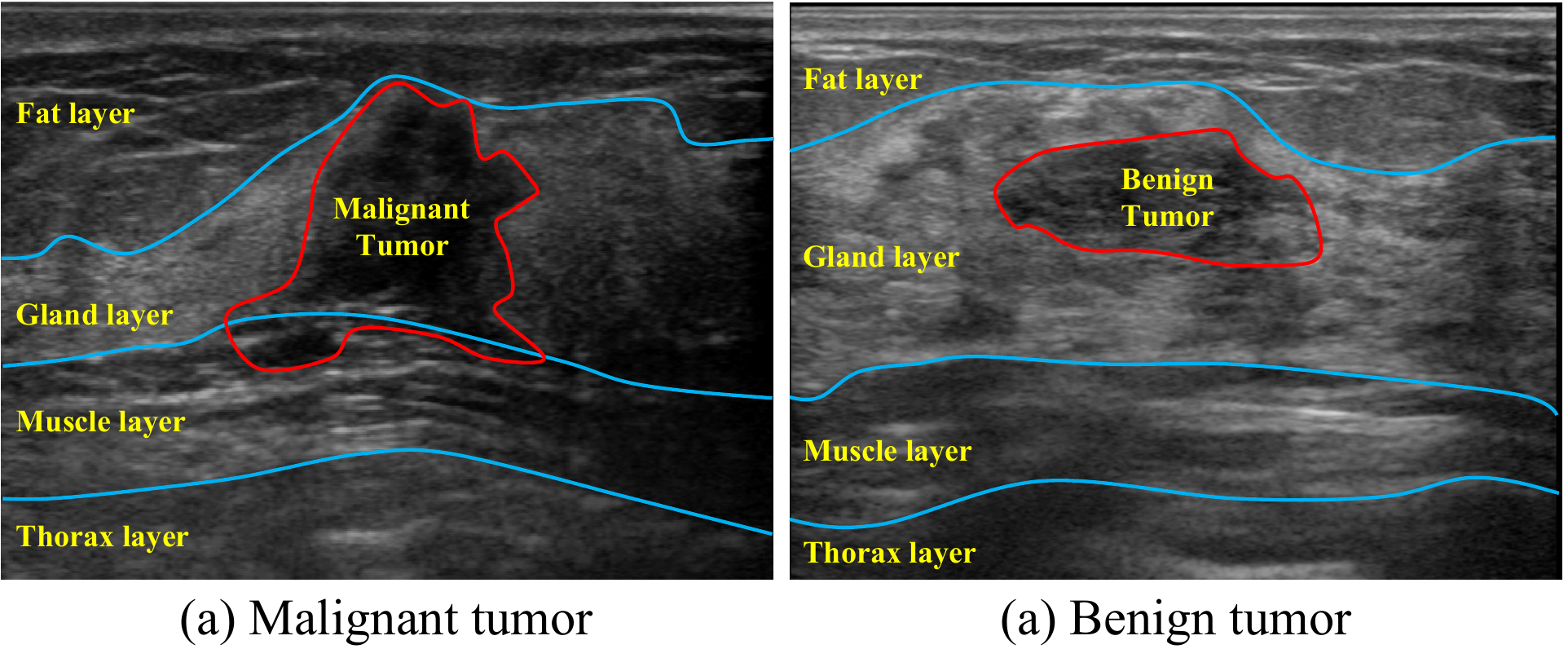}
	\caption{Anatomical structure of breast in ultrasound images. (a) Malignant tumor. (b) Benign tumor.}
	\label{fig:structure}
\end{figure}

In this paper, we promote automatic breast cancer diagnosis in ultrasound images in the following two aspects, data resources and methodology. First, we release a large breast lesion classification dataset \textit{GDPH\&SYSUCC}, which was collected from two medical centers with 886 benign and 1519 malignant images, for a total of 2405 BUS images. We only provide the whole BUS images without ROI annotations. Then, we propose an interpretable breast cancer diagnosis model for BUS. We find that the sequential data analysis nature of the transformer perfectly fits the anatomical prior of the breast ultrasound image. As shown in Fig.~\ref{fig:structure}, there are four layers from top to bottom, including subcutaneous fat layer, gland layer, muscle layer and thorax layer. Malignant tumors always start from the gland layer and invade the deeper layers. Benign breast tumors typically originated in the glandular tissue and destructed the gland continuity. Therefore, we design an anatomy-aware model, called HoVer-Transformer (in short HoVer-Trans), which considers the prior knowledge of the anatomical structure in BUS. We propose a HoVer-Trans block to extract the inter-layer spatial information \textit{horizontally} and the intra-layer spatial information \textit{vertically}. In HoVer-Trans, we introduce convolutional layers to joint two adjacent transformer stages to fuse the horizontal and vertical image features and to introduce inductive bias.

The proposed HoVer-Trans is evaluated by extensive experiments, including comparisons with SOTA methods in several datasets, model interpretability and ablation studies. HoVer-Trans achieves comparable quantitative performance in all the datasets. The visualization heatmaps also demonstrate that HoVer-Trans is able to pay attention to the malignant lesion boundary (invasive margin), which proves the horizontal and vertical design successfully learned anatomical prior knowledge. Ablation studies demonstrate the effectiveness of each specific technical design. We also compare our proposed model with two senior sonographers. HoVer-Trans outperforms both sonographers in the entire dataset and BI-RADS subgroup analysis under the same condition. The main contributions of this paper are summarized as follows.

\begin{itemize}
  \item[$\bullet$] We release a new breast cancer classification dataset, GDPH\&SYSUCC which is the largest open dataset in this field.
  \item[$\bullet$] We propose an anatomy-aware model HoVer-Trans to fully automatically classify the breast lesion and achieve comparable performance compared with six baseline models.
  \item[$\bullet$] HoVer-Trans is able to provide the interpretable evidence to support the decision of the model.
\end{itemize}

\section{Related Works}
Deep learning techniques have dominated almost all the medical imaging modalities and tasks, including histopathology image segmentation~\cite{graham2019hover}, MRI~\cite{liu2020ms}, CT~\cite{li2018h} and etc. In this section, we summarize the previous researches on breast cancer diagnosis in ultrasound images~\cite{ilesanmi2021methods,jimenez2020deep} and the transformer-based medical image classification models~\cite{liu2021survey}.

\subsection{Breast Cancer Diagnosis in Ultrasound Images}
Ultrasonography is one of the most common and non-invasive imaging modalities for breast cancer screening and diagnosis. Precisely detecting and diagnosing malignant tumors allows early intervention to reduce mortality. It generally relies on the subjective evaluation of the sonographers. However, manual assessment highly depends on the clinical experience due to the heterogeneous of the malignant tumors and the low image quality of the ultrasound images. Therefore, it is essential to design CAD algorithms to automatically and objectively evaluate breast ultrasonography. With the recent advances in artificial intelligence techniques, like radiomics~\cite{gillies2016radiomics} and deep learning~\cite{lecun2015deep}, researchers started to solve various clinical prediction applications in a data-driven manner and achieved outstanding performances in breast cancer diagnosis, such as lesion classification~\cite{ning2020multi,qian2021prospective}, axillary lymph node status prediction~\cite{zheng2020deep,zhou2020lymph}, sentinel lymph node status prediction~\cite{guo2020deep,zha2021preoperative} and even molecular status prediction~\cite{xu2022predicting,jiang2021deep}.

Currently, deep learning-based models have already dominated the breast lesion classification. Flores~\etal~\cite{flores2015improving} explored the prediction value of the morphological and texture features for breast lesion classification on ultrasonography.
Byra~\etal~\cite{byra2019breast} transferred the model pre-trained on ImageNet to fine-tune the breast mass classification model, which is a popular way for small- or mid-sized data.
Some researchers~\cite{erouglu2021convolutional,moon2020computer} attempted to ensemble the deep features from multiple classification architectures and applied machine learning classifiers for breast ultrasonography image classification.
Zhuang~\etal~\cite{zhuang2020breast} proposed an image decomposition and enhancement method to enrich the information of the ultrasound image.
Qian~\etal~\cite{qian2021prospective} aggregated the multimodal ultrasound images for an explainable prediction to support the clinical decision-making of the sonographers and increase the confidence levels of the decision.
Cui~\etal~\cite{cui2022fmrnet} proposed an FMRNet to fuse combined tumoral, intratumoral and peritumoral regions to represent the whole tumor heterogeneous.
Di~\etal~\cite{di2022saliency} introduced a saliency-guided approach to differentiate the foreground and background regions by two separated branches. A hierarchical feature aggregation branch was proposed to fuse the features from both branches and make the inference.
Qi~\etal~\cite{qi2019automated} designed two identical CNN backbones to identify the malignant tumor and solid nodule separately. The class activation maps generated from two CNN backbones were used to guide each other. They validated the proposed model in a large dataset with 8145 breast ultrasonography images. Unfortunately, the dataset in this paper is private.
Shen~\etal~\cite{shen2021artificial} discussed how the CAD system helps sonographers reduce the false positive rate.

For the breast lesion classification, even though existing models have already achieved comparable performance with sonographers, it is still worth to keep discovering the potential values of deep learning models from different perspectives. People now pay more attention to clinical usability other than only considering the accuracy. The clinical usability reflects in the following factors. (1) \textit{Accuracy}: Whether the model prediction results are more accurate than sonographers. (2) \textit{Interpretability}: Whether the model can provide any sonographic symptom or evidence to support the decision. (3) \textit{Model Convenience}: Whether the model is fully automated without any user input, like manual segmentation or predefined ROIs.


\begin{figure*}[t]
	\includegraphics[width=.99\linewidth]{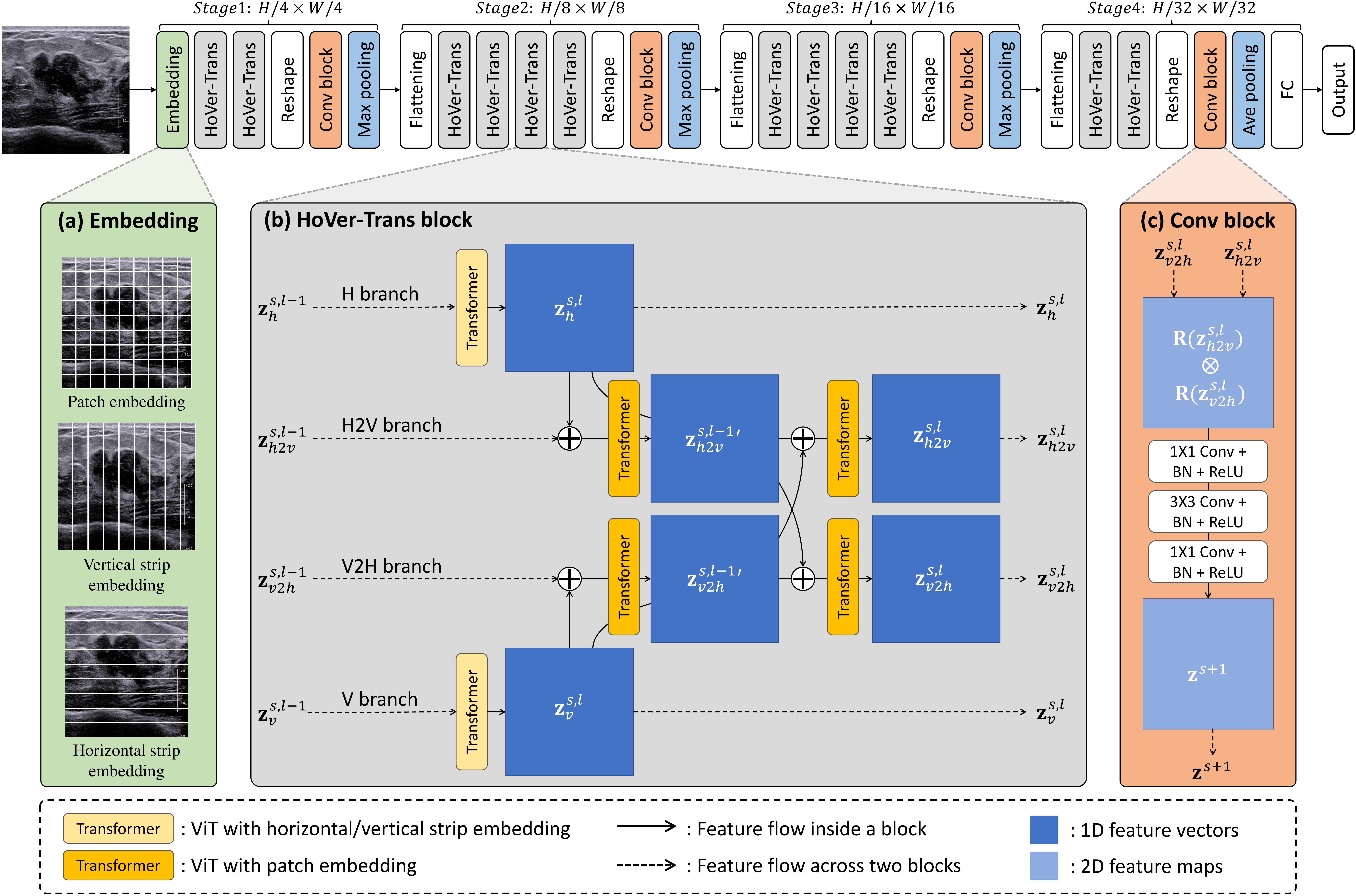}
	\caption{Network architecture of the proposed model. It contains four stages, each of which consists of several HoVer-Trans blocks, a convolutional block and a pooling layer. (a) Three embedding ways of patch embedding, vertical strip embedding and horizontal strip embedding. (b) HoVer-Trans formulates the anatomical prior knowledge in breast ultrasound images, which is designed to extract the intra-layer and inter-layer relationships of the anatomical layers in the breast. It consists of four branches. The horizontal branch and the vertical branch are designed to extract the inter-layer and intra-layer relationships respectively. H2V and V2H branches are introduced to fuse the horizontal and vertical features. The output features from each branch in the HoVer-Trans block will be regarded as the input features of the next HoVer-Trans block. (c) Conv block is applied to connect two stages and to introduce inductive bias.}
	\label{fig:network}
\end{figure*}

\subsection{Transformer-based Medical Image Classification}
Transformer~\cite{vaswani2017attention} is originally designed for natural language processing. It has been widely used in sequential data analysis thanks to the elegant self-attention mechanism~\cite{lin2021survey}. The invention of the vision transformer (ViT)~\cite{dosovitskiy2020image} is leading the transformer-based models toward the computer vision applications by cropping the image into several small tiles (visual words). Swim transformer~\cite{liu2021swin} introduces multi-scale information like a CNN model does by a hierarchical structure and shifted windows. Sooner, various transformer models~\cite{shamshad2022transformers} were proposed for medical image classification, such as COVID-VIT for CT chest COVID-19 classification~\cite{gao2021covid}, TransMIL for pathology image classification\cite{shao2021transmil}, MIL-VT for fundus image classification~\cite{yu2021mil} and etc. 

Instead of designing a complex black-box model for breast cancer prediction, we intend to take model interpretability, generalizability and convenience into consideration. By formulating the anatomical prior knowledge into the transformer model design, the proposed model demonstrates superior predictive ability while can provide interpretable features to support the decisions.

\section{Methodology}
Since the anatomical structures of the breast are very clear in the ultrasound images. By leveraging this prior knowledge, we propose an anatomy-aware model for fully automatic breast cancer diagnosis in ultrasound images. In this section, we demonstrate the methodology of the proposed model, shown in Fig.~\ref{fig:network}. First, we introduce the key idea of the anatomy-aware formulation in Sec.~\ref{sec:method_idea}. Based on this idea, the HoVer-Trans stage is proposed in Sec.~\ref{sec:method_hover}. Next, we define the overall network structure of the proposed model in Sec.~\ref{sec:method_overall}. Sec.~\ref{sec:implement} shows the implementation details.

\subsection{Anatomy-aware Formulation}\label{sec:method_idea}

\begin{figure}
	\includegraphics[width=.99\linewidth]{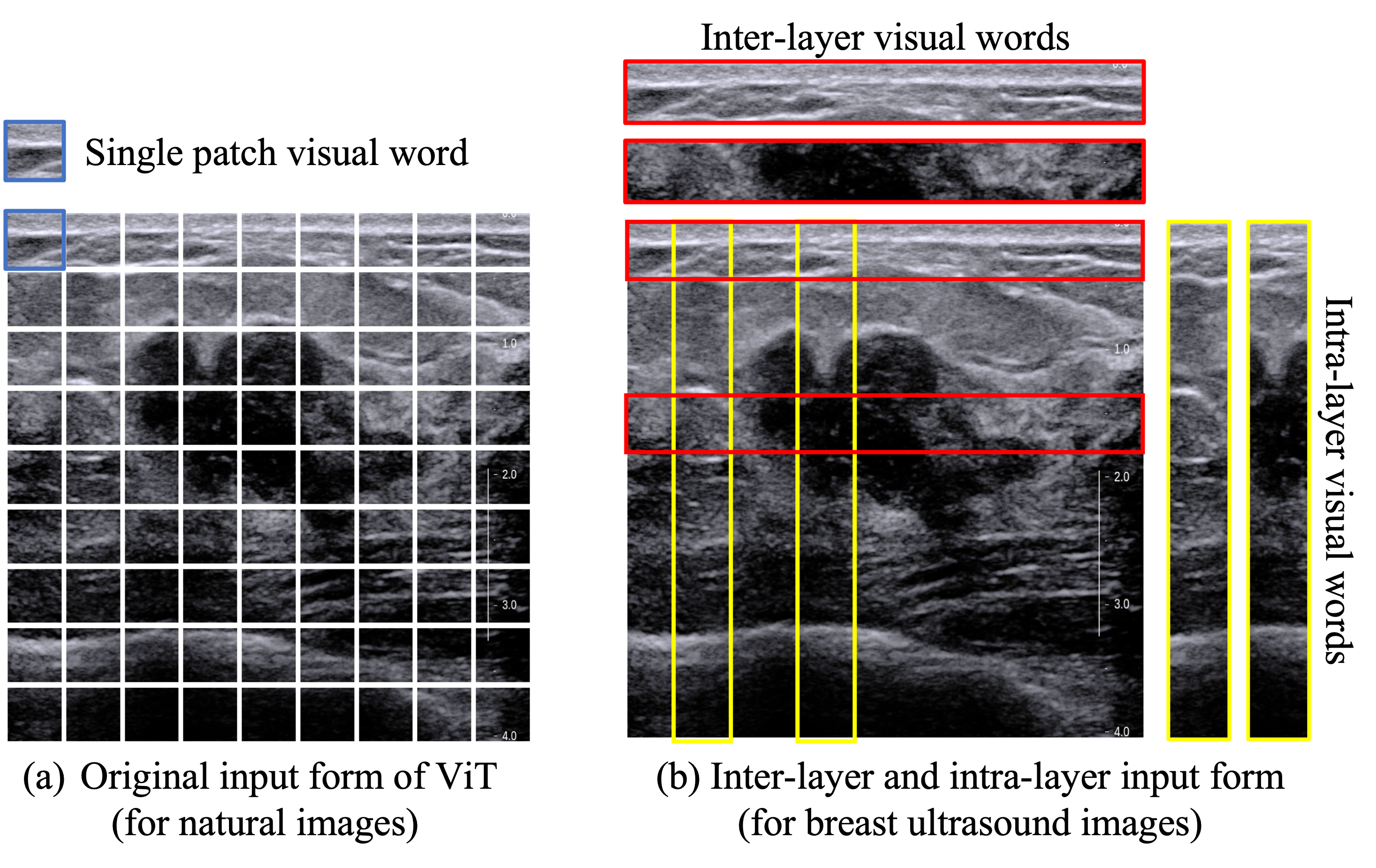}
	\caption{Comparison of two embedding ways. (a) ViT tessellates the image into several 16$\times$16 patches (visual words). (b) We formulate the anatomical structure of breast by the inter-layer visual words (horizontal strips) and the intra-layer visual words (vertical strips).}
	\label{fig:idea}
\end{figure}
According to the ultrasound imaging principles and the anatomical structure of the breast, different breast tissues form different layers clearly in the ultrasound images, as shown in Fig.~\ref{fig:structure}. The size, location and morphological appearance of the lesion and the spatial relationship with different layers determine the malignancy of the lesion. Conventional CNN models are good at extracting representative local features but show less effective spatial relationship representation ability. That is the reason why most of the existing breast cancer diagnosis algorithms in ultrasound images need a pre-defined ROI of the lesion to remove the redundant area and let the CNN model classify the ROI. The self-attention nature of the transformer introduces strong spatial relationships of each visual word, as shown in Fig.~\ref{fig:idea}~(a). To further exploit the intra-layer and inter-layer spatial correlations in BUS, we formulate the problem by transforming the square-shape visual words into horizontal and vertical strips to bring the anatomical prior knowledge into the model, as shown in Fig.~\ref{fig:idea}~(b). 

\subsection{HoVer-Trans Stage}\label{sec:method_hover}
Since our proposed model is constructed on top of the ViT structure, we first briefly introduce ViT at the beginning of this part. And then we show how the proposed HoVer-Trans stage is constructed.

\subsubsection{Vision Transformer}
Vision transformer (ViT)\cite{dosovitskiy2020image} is the first to bring the most popular technique in natural language processing into the computer vision world. It tessellates the input image $\textbf{x} \in \mathbb{R}^{H\times W\times C}$ into patches $\textbf{x}_p \in \mathbb{R}^{N\times (P^2 \cdot C)}$ and regards them as the visual words (tokens), where $(H,W,C)$ and $(P,P,C)$ are the resolution with channels of the input image and the patches, respectively. $N$ is the number of patches. For each visual word, they transform the 2D image into a 1D vector, called patch embedding. Multi-head self-attention mechanism builds spatial correlations across different tokens. The formulation of ViT is shown as follows:

\begin{equation}
\begin{split}
&\textbf{z}_0 = [\textbf{x}_{class}; \textbf{x}_p^1\textbf{E}, \textbf{x}_p^2\textbf{E},..., \textbf{x}_p^N\textbf{E},] + \textbf{E}_{pos},\\
& \quad \textbf{E}\in \mathbb{R}^{(P^2\cdot C)\times D}, \textbf{E}_{pos}\in \mathbb{R}^{(N+1)\times D}
\end{split}
\end{equation}

\begin{equation}
	\textbf{z}'_l = {\rm MSA}({\rm LN}(\textbf{z}_{l-1}))+\textbf{z}_{l-1}, \quad l = 1...L
\end{equation}

\begin{equation}
	\textbf{z}_l = {\rm MLP}({\rm LN}(\textbf{z}'_l))+\textbf{z}'_l, \quad l = 1...L
\end{equation}

\begin{equation}
	\textbf{y}={\rm LN}(\textbf{z}_L^0)
\end{equation}

In our proposed HoVer-Trans, we use several ViT blocks with exactly the same structure without class embedding to construct the HoVer-Trans block. Thus, we denote all the ViT blocks in the following paper as $\texttt{Trans}(\cdot)$.

\subsubsection{Embedding}
To formulate the anatomical prior knowledge into the transformer model, we introduce additional two embedding ways shown in Fig.~\ref{fig:network}~(a), following the idea presented in Sec.~\ref{sec:method_idea}. Given an input BUS image $I \in \mathbb{R}^{H\times W\times C}$, patch embedding, horizontal strip embedding and vertical strip embedding are processed before feeding them into the model. The patching embedding cuts the input image into $N\times N$ patches $\textbf{x}_p^{(r,c)}$, where $r$ and $c$ denote the indices of the row and column. After flattening and linear projection, we get a group of 1D vectors $\textbf{z}_{p}$.

\begin{equation}\label{eq:p_embedding}
	\textbf{z}_{p}=\{\textbf{x}_{p}^{(r,c)}|\textbf{x}\in \mathbb{R}^{{H/N}\times {W/N} \times C}\}, \quad r, c=1, ..., N
\end{equation}

Horizontal strip embedding is introduced to represent the visual words of the same anatomical layer with $M$ strips, defined as follows:
\begin{equation}\label{eq:h_embedding}
	\textbf{z}_{h}=\{\textbf{x}_h^{(m)}|\textbf{x}\in \mathbb{R}^{{H/M}\times W \times C}\}, \quad r=1, ..., M
\end{equation} 

Vertical strip embedding is introduced to represent the visual words across anatomical layers with $M$ strips, defined as follows:
\begin{equation}\label{eq:v_embedding}
	\textbf{z}_{v}=\{\textbf{x}_v^{(m)}|\textbf{x}\in \mathbb{R}^{H\times {W/M} \times C}\}, \quad c=1, ..., M
\end{equation} 

\subsubsection{HoVer-Trans Block}
The architecture of the HoVer-Trans block is depicted in Fig.~\ref{fig:network}~(b). We design a symmetry structure with four branches in one HoVer-Trans block, H branch (horizontal), V branch (vertical), H2V branch (horizontal to vertical) and V2H branch (vertical to horizontal). 

Let us define the features at the $l$-th block in the $s$-th stage as $\textbf{z}^{s,l}_{\{h,v,h2v,v2h\}}$. HoVer-Trans block takes the outputs from the previous block and generates the features for the next block, defined as follows. 
\begin{equation}
	\{\textbf{z}_{h}^{s,l},\textbf{z}_{v}^{s,l},\textbf{z}_{h2v}^{s,l},\textbf{z}_{v2h}^{s,l}\}= f(\textbf{z}_{h}^{s,l-1},\textbf{z}_{v}^{s,l-1},\textbf{z}_{h2v}^{s,l-1},\textbf{z}_{v2h}^{s,l-1})
\end{equation}
Where $f(\cdot)$ denotes the HoVer-Trans block. The inputs of four branches in the first HoVer-Trans block (when $l=1$) are equivalent to the features from the previous HoVer-Trans stage $\textbf{z}^{s-1}$.
\begin{equation}
	\textbf{z}^{s,1}_{\{h,v,h2v,v2h\}} = \textbf{z}^{s-1}
\end{equation}

\textbf{H and V branches} are two auxiliary branches to extract the inter-layer and intra-layer spatial correlations, with two identical H and V branches with horizontal strip embedding (defined in Eq.~\ref{eq:h_embedding}) and vertical strip embedding (defined in Eq.~\ref{eq:v_embedding}), respectively. 
\begin{equation}
	\textbf{z}_{h}^{s,l}=\texttt{Trans}(\textbf{z}_{h}^{s,l-1})
\end{equation}

\begin{equation}
	\textbf{z}_{v}^{s,l}=\texttt{Trans}(\textbf{z}_{v}^{s,l-1})
\end{equation}

The anatomy-aware spatial features $\textbf{z}_{h}^{s,l}$ and $\textbf{z}_{v}^{s,l}$ will be passed into the next two main branches (H2V and V2H). 
They are also regarded as the inputs of the next HoVer-Trans block.

\textbf{H2V and V2H branches} are served as the main feature extraction branches which fuse the features from two auxiliary branches (H and V). For example in the H2V branch, the horizontal features $\textbf{z}_{h}^{s,l}$ are added to the features from the previous HoVer-Trans block $\textbf{z}_{h2v}^{s,l-1}$. After a transformer encoder, the vertical features $\textbf{z}_{v}^{s,l}$ are added behind. The V2H branch is the mirror of the H2V branch.
\begin{equation}
	\textbf{z}_{h2v}^{s,l}=\texttt{Trans}(\texttt{Trans}(\textbf{z}_{h}^{s,l}+\textbf{z}_{h2v}^{s,l-1})+\textbf{z}_{v}^{s,l})
\end{equation}

\begin{equation}
	\textbf{z}_{v2h}^{s,l}=\texttt{Trans}(\texttt{Trans}(\textbf{z}_{v}^{s,l}+\textbf{z}_{v2h}^{s,l-1})+\textbf{z}_{h}^{s,l})
\end{equation}

The output features $\textbf{z}_{h2v}^{s,l}$ and $\textbf{z}_{v2h}^{s,l}$ will be passed into the next block. Note that, for the last HoVer-Trans Block in each stage, the features will be passed into a Conv Block, described as follows.

\subsubsection{Conv Block}
The transformer is good at processing sequential data and extracting spatial correlations. But it lacks inductive bias. In order to leverage the strength of both transformer and CNN, we introduce a convolutional block (Fig.~\ref{fig:network}~(c)) after the last HoVer-Trans block at each stage to fuse the H2V and V2H features and introduce the inductive bias. The Conv block consists of three convolutional layers. In the first convolutional layer with $1\times 1$ kernel size, we set the number of output channels to be twice the number of input channels. In the second convolutional layer, a $3\times 3$ convolution kernel is used. In the third convolutional layer $1\times 1$, the number of channels is compressed to adapt to the next stage.

The features from two main branches (H2V and V2H) are reshaped to 2D feature maps and concatenated together before being processed by the convolutional block.
\begin{equation}
	\textbf{z}^{s+1}=\texttt{Conv}(\texttt{Concat}[\texttt{R}(\textbf{z}_{h2v}^{s,l}),\texttt{R}(\textbf{z}_{v2h}^{s,l})])
\end{equation} 
where $\texttt{R}(\cdot)$ means to reshape the 1D feature vectors to 2D feature maps.

\subsection{Overall Architecture}\label{sec:method_overall}
The overall structure of our model is shown in Fig~\ref{fig:network}. The model consists of four $stage$ modules. Each stage module consists of several HoVer-Trans blocks, one Conv block and one pooling layer.

Given a BUS image $I \in \mathbb{R}^{H\times W\times 3}$, we first use convolutional stem~\cite{NEURIPS2021_convstem} to downscale $I$ the image size to $H/4\times W/4\times C$. The resolution of the feature maps in the next three stages are $H/8\times W/8\times 2C$, $H/16\times W/16\times 4C$, $H/32\times W/32\times 8C$, which is similar to the structure of the traditional convolutional neural network~\cite{he2016deep,vgg}. To fuse the horizontal and vertical information, a Conv block is introduced to connect two adjacent stages. So the input of each stage is a 2D image or 2D feature maps. Embedding or flattening will be introduced to fit the input of the transformer. In the last stage, the fully connected layer is applied for inference.
The model is optimized by the cross-entropy loss. 

\begin{figure}[t]
	\includegraphics[width=.99\linewidth]{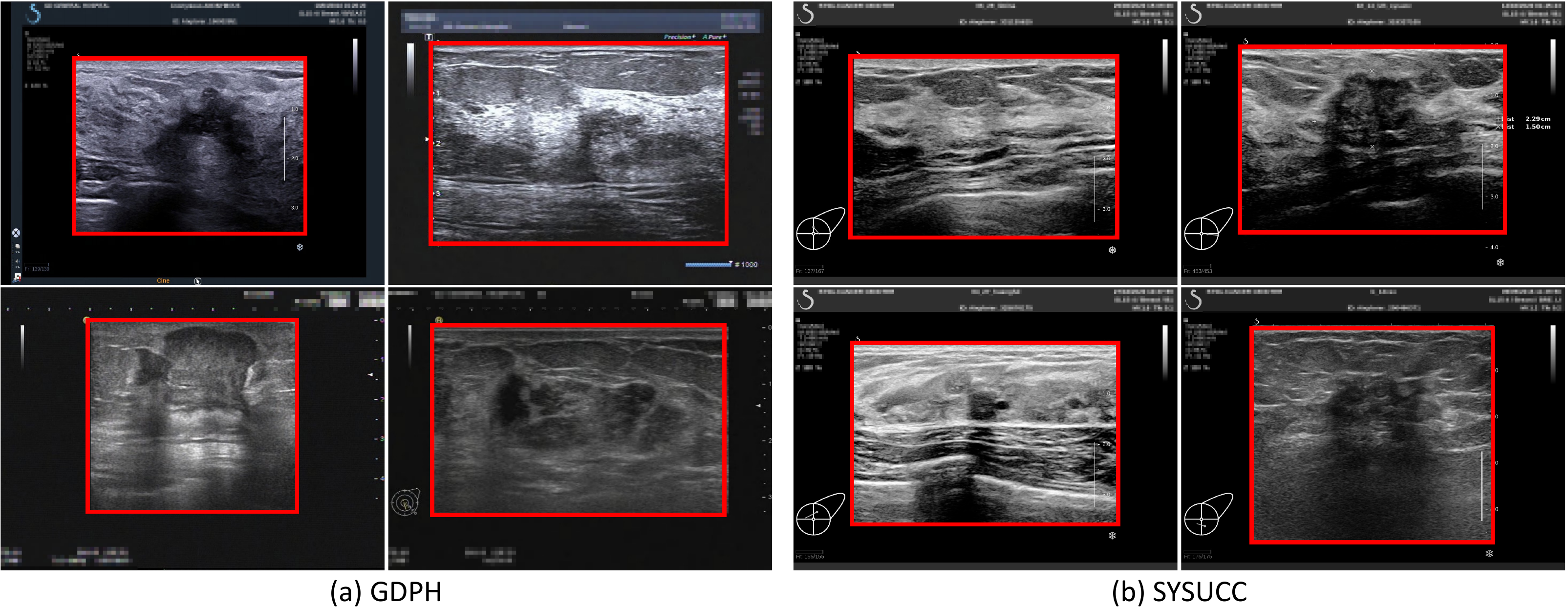}
	\caption{Some examples of breast ultrasound images in two medical centers, (a) GDPH and (b) SYSUCC. The red boxes are the foreground images we extract.}
	\label{fig:BUS example}
\end{figure}

\subsection{Implementation Details}\label{sec:implement}
We use Python3.6 and PyTorch1.8 to implement all the models. All the experiments are run with an 11 GB NVIDIA GeForce RTX 2080Ti GPU. We build our model with embedding dimensions of each stage of $\{4, 8, 16, 32\}$, and HoVer-Trans block numbers of each stage are $\{2,4,4,2\}$. Head numbers of transformer block in each stage are $\{2,4,8,16\}$. We train for 250 epochs with the AdamW~\cite{Adam} optimizer, a batch size of 32, weight decay of 0.1, 10 warm-up epochs and an initial learning rate of 0.0001 with a cosine decay learning rate scheduler. The augmentation strategy includes blurring, noise, horizontal flip, brightness and contrast. Because the order of the tissue layers is fixed, we do not use vertical flip data augmentation. All the images will be resized to $256\times 256$.
\begin{table}[t]
	\centering
	\caption{Distribution of mass according to BI-RADS stratification. BI-RADS stratification are categorized as into 6 classes. Class 2 to 5, including 3 sub-categories 4A, 4B and 4C.}
	\label{tab:distribution}
	\begin{tabular}{c|c|c|c}
		\hline
		BI-RADS & Benign & Malignant & Total\\
		\hline
		2 &66  &0   &66\\
		3 &610 &101 &711\\
		4A&162 &659 &821\\
		4B&38  &495 &533\\
		4C&10  &193 &203\\
		5 &0   &71  &71\\
		\hline
		Total&886&1519&2405\\
		\hline
	\end{tabular}
\end{table}
\section{Datasets}\label{sec4}
In this paper, we use three datasets to evaluate the diagnostic performance of our model, two of which are the public datasets and one is our constructed dataset. 

\subsection{UDIAT}
The first public dataset is a small dataset, named UDIAT~\cite{yap2017automated}, which contains a total of 163 BUS images with 109 BUS images of benign lesions and 54 BUS images of malignant lesions. All the images were collected from the UDIAT Diagnostic Centre of the Parc Tauli Corporation, Sabadell, Spain. The average size of the images is $760\times570$ pixels and the range from $307\times 233$ to $791\times641$.

\begin{table*}[t]
	\centering
	\caption{Quantitative comparisons with SOTA approaches in three datasets. The last column shows the p-value of DeLong's test between the AUC of each baseline model and HoVer-Trans model. P-values less than 0.05 are marked as *, p-values less than 0.01 are marked as **,  and p-values less than 0.001 are marked as ***. Model with $\dagger$ means they require a pre-defined ROI.}
	\label{tab:quantitative}
	\begin{tabular}{c|cccccc|c}
		\toprule[2pt]
		\multicolumn{8}{c}{UDIAT}\\
		\hline
		& AUC & ACC & Specificity & Precision & Recall & F1-score& $p$-value\\
		\hline
		ResNet50    &0.778$\pm$0.059 &0.743$\pm$0.073 &\textbf{0.899$\pm$0.118} &0.676$\pm$0.146 &0.426$\pm$0.256 &0.523$\pm$0.120 &***\\
		VGG16       &\textbf{0.786$\pm$0.073} &0.756$\pm$0.123 &0.800$\pm$0.106 &0.650$\pm$0.120 &\textbf{0.672$\pm$0.183} &\textbf{0.661$\pm$0.107} &***\\
		ViT         &0.740$\pm$0.140 &0.701$\pm$0.300 &0.880$\pm$0.147 &0.606$\pm$0.240 &0.364$\pm$0.132 &0.455$\pm$0.223 &***\\
		TNT-s       &0.627$\pm$0.082 &0.626$\pm$0.089 &0.752$\pm$0.150 &0.426$\pm$0.276 &0.370$\pm$0.241 &0.396$\pm$0.160 &***\\
		Ours		&0.781$\pm$0.118 &\textbf{0.774$\pm$0.061} &0.889$\pm$0.128 &\textbf{0.714$\pm$0.214} &0.545$\pm$0.232 &0.619$\pm$0.099 &-\\
		\hline
		MICCAI2020$\dagger$&0.939$\pm$0.031&0.909$\pm$0.032&0.927$\pm$0.106&0.900$\pm$0.044&-&-&-\\
		JBHI2020$\dagger$&0.870&0.859&0.685&0.945&0.840&-&-\\
		\toprule[2pt]
		
		\multicolumn{8}{c}{BUSI}\\
		\hline
		ResNet50	&0.877$\pm$0.034 &0.818$\pm$0.039 &0.883$\pm$0.030 &0.738$\pm$0.049 &0.682$\pm$0.079 &0.709$\pm$0.062 &***\\
		VGG16       &\textbf{0.898$\pm$0.037} &0.832$\pm$0.041 &0.778$\pm$0.056 &0.873$\pm$0.054 &0.862$\pm$0.096 &0.867$\pm$0.063 &***\\
		ViT 		&0.834$\pm$0.062 &0.811$\pm$0.052 &\textbf{0.922$\pm$0.070} &0.781$\pm$0.146 &0.579$\pm$0.032 &0.665$\pm$0.094 &***\\
		TNT-s		&0.852$\pm$0.015 &0.812$\pm$0.032 &0.908$\pm$0.035 &0.763$\pm$0.057 &0.611$\pm$0.104 &0.679$\pm$0.057 &***\\
		Ours 		&0.865$\pm$0.066 &\textbf{0.855$\pm$0.050} &0.838$\pm$0.053 &\textbf{0.876$\pm$0.062} &\textbf{0.867$\pm$0.115} &\textbf{0.872$\pm$0.080} &-\\
		\hline
		JBHI2020$\dagger$&0.889&0.843&0.758&0.883&0.751&-&-\\
		
		\toprule[2pt]
		\multicolumn{8}{c}{GDPH\&SYSUCC}\\
		\hline
		ResNet50    &0.886$\pm$0.014 &0.832$\pm$0.014 &0.732$\pm$0.033 &0.851$\pm$0.015 &0.890$\pm$0.013 &0.870$\pm$0.010 &**\\
		VGG16       &0.919$\pm$0.006 &0.864$\pm$0.004 &\textbf{0.892$\pm$0.010} &0.811$\pm$0.009 &0.814$\pm$0.007 &0.813$\pm$0.003 &*\\
		ViT         &0.806$\pm$0.021 &0.734$\pm$0.029 &0.694$\pm$0.053 &0.809$\pm$0.047 &0.758$\pm$0.021 &0.782$\pm$0.028 &***\\
		TNT-s       &0.853$\pm$0.010 &0.781$\pm$0.015 &0.618$\pm$0.059 &0.793$\pm$0.050 &0.879$\pm$0.028 &0.834$\pm$0.015 &***\\
		Ours        &\textbf{0.924$\pm$0.016} &\textbf{0.893$\pm$0.021} &0.836$\pm$0.038 &\textbf{0.906$\pm$0.023} &\textbf{0.926$\pm$0.019} &\textbf{0.916$\pm$0.019} &-\\
		\hline
		JBHI2020$\dagger$&0.856$\pm$0.009&0.811$\pm$0.022&0.859$\pm$0.038&0.824$\pm$0.022&0.891$\pm$0.028&0.856$\pm$0.020 &***\\
		\toprule[2pt]
		
	\end{tabular}
\end{table*}

\subsection{BUSI}
The second dataset, BUSI~\cite{al2020dataset}, consists of a total number of 780 BUS images from the Baheya Hospital for Early Detection and Treatment of Women’s Cancer, Cairo, Egypt. BUSI dataset includes 210 images with benign lesions, 437 images with malignant lesions, and 133 normal BUS images without lesions. Furthermore, each image has the pixel-level ground truth of the lesion. In this paper, since we only differentiate the malignant and benign lesions, normal BUS images are excluded. 647 images are finally utilized with the average resolution of $608\times494$ pixels and the range from $190\times335$ to $916\times683$.

\subsection{GDPH\&SYSUCC}
In this study, we also construct a publicly available dataset of BUS images for breast cancer diagnosis. The BUS images came from two medical centers. 1) the Department of Ultrasound, Guangdong Provincial People’s Hospital, Guangzhou, Guangdong, China (GDPH). 2) the Department of Ultrasound, Sun Yat-sen University Cancer Center, Guangzhou, Guangdong, China (SYSUCC). We exported the images and their corresponding BI-RADS scores from the picture archiving and communication system (PACS). The ultrasound images were acquired from following equipments, including Hitachi Ascendus (Japan), Mindray DC-80 (China), Toshiba Aplio 500 (Japan) and Supersonic Aixplorer (France). All the images were labeled as benign or malignant according to the pathology report after the biopsy or surgery was performed. The dataset consists of 1519 malignant BUS images from 676 patients and 886 benign BUS images from 526 patients, for a total of 2405 BUS images. The average size of the images is $844\times627$ and the range from $278\times 215$ to $1280\times800$. Fig.~\ref{fig:BUS example} shows some examples of BUS images in our dataset. To protect the privacy of the patients, we mosaic the personal information. The distribution of the BI-RADS scores and the statistics of the malignant tumors and the benign tumors are shown in Table~\ref{tab:distribution}.

\subsubsection{Data Preprocessing}
All the images are exported from the picture archiving and communication system (PACS). We extract the image data from the original DICOM format files of BUS and save them in PNG format. To exclude the UI regions, we extract the foreground images by applying a rectangle detection algorithm provided by the OpenCV library, and manually check all the images to ensure the completeness of the foreground images.
\section{Experimental Results}\label{sec5}
In this section, we first demonstrate the experimental setting in Sec.~\ref{sec:exp_setting}, including the evaluation metrics and the competitors. Then we show the quantitative results and the heatmap visualizations of three datasets in Sec.~\ref{sec:exp_result}. In Sec.~\ref{sec:AIvsHuman}, we compare our proposed model with two senior sonographers. Ablation studies have been conducted in Sec.~\ref{sec:exp_ablation}. Heatmap visualizations of the model are shown in all the experiments to evaluate the interpretability.

\subsection{Experimental Setting}\label{sec:exp_setting}
\subsubsection{Evaluation Metrics}
To comprehensively evaluate the performance of the proposed model, we introduce the following metrics, including area under the ROC curve (AUC), accuracy (ACC), specificity, precision, recall and F1 score.

\subsubsection{Competitors}
In this paper, we compare our proposed model with six state-of-the-art (SOTA) models, including two most popular CNN-based classification models ResNet50~\cite{he2016deep} and VGG16~\cite{vgg}, two vision transformer models ViT~\cite{dosovitskiy2020image} and TNT-s~\cite{han2021transformer} and two CNN-based models tailored for breast cancer diagnosis in ultrasound images JBHI2020~\cite{xing2020using} and MICCAI2020~\cite{ning2020multi}. Although JBHI2020 and MICCAI2020 are both designed for breast cancer diagnosis, they are still a little bit different from our proposed model. Both JBHI2020 and MICCAI2020 do not process the entire ultrasound images. These two models have to first pre-define a region of interest (ROI) of the mass, and then classify the malignancy of the corresponding lesion. Furthermore, JBHI2020 includes additional BI-RADS scores in the training phase. So in the existing two public datasets UDIAT and BUSI, the quantitative performance of these two models is directly copied from the corresponding papers. In the GDPH\&SYSUCC dataset, we compare our proposed model with the other four image classification baseline models without ROIs of the lesions. We also implement the model from JBHI2020~\cite{xing2020using} and test it in the GDPH\&SYSUCC dataset. Since JBHI2020 requires the pre-defined ROI of the lesion, we invited a sonographer to label the bounding box of each lesion for this approach.

\begin{figure*}[htbp]
	\includegraphics[width=.99\linewidth]{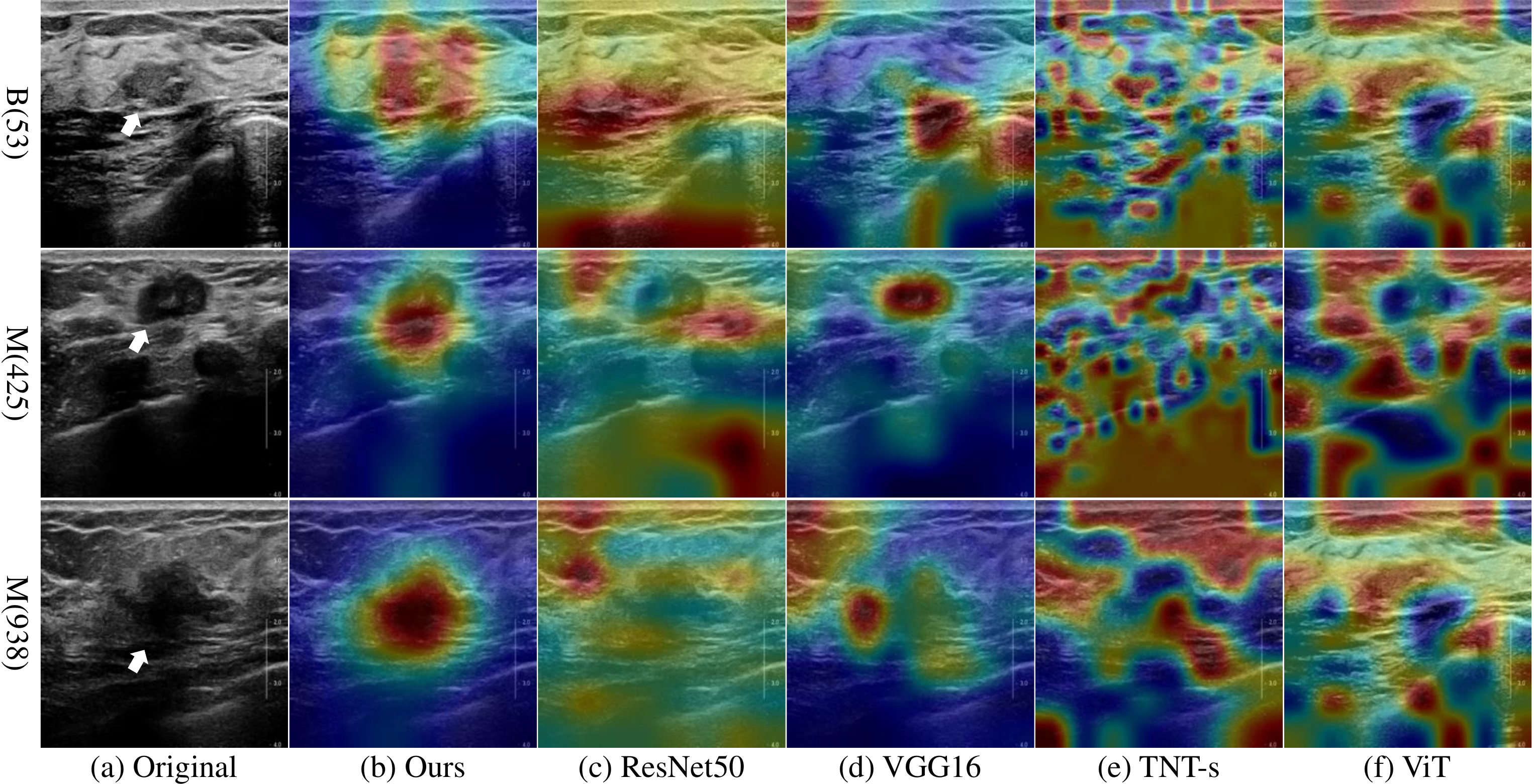}
	\caption{The heatmaps of different models (GDPH\&SYSUCC). We overlay the heatmaps generated by feature maps on the original BUS images. (a) are the original images. (b)-(f) are the heatmap visualizations of different models. M and B indicate the images with malignant tumors and benign tumors, respectively. The lesions are pointed by the white arrows.}
	\label{fig:fm_compare}
\end{figure*}

\subsection{Comparisons with SOTA Approaches}\label{sec:exp_result}
\subsubsection{Performance of deep-learning models}
Table~\ref{tab:quantitative} shows the quantitative results in three datasets. In the UDIAT dataset, since MICCAI2020 and JBHI2020 classify the lesion within the ROIs, it can alleviate the underfitting problem when the dataset only contains 163 images by removing the regions without lesions. Among the other five models, VGG16 and our model achieve promising classification results even with such a small dataset. In the BUSI dataset (647 images), the classification performance of our proposed model achieves the best ACC of 0.855, the precision of 0.876, the recall of 0.867 and the F1-score of 0.872. A larger dataset with more training samples allows the neural network models to learn better feature representation. The performance of our model in the BUSI dataset is comparable to the performance of the ROI-based model JBHI2020.

Most of the existing approaches, including MICCAI2020 and JBHI2020, solve the BUS classification problem by a two-step approach. Pre-defined ROIs can remove the regions without tumors, which might be good for the small dataset. However, when we have enough data, it is hard to say whether removing these regions do more good than harm. Because it also loses the spatial information of the tumor in the breast, which is also an important clue for breast cancer diagnosis. In our dataset GDPH\&SYSUCC (2405 images), the proposed HoVer-Trans achieves the best classification performance on AUC, ACC, precision, recall and F1-score. It outperforms all the baseline models, including JBHI2020. When with enough training data, the advantage of the anatomy-aware design is demonstrated. Considering both the tumor region and the surrounding area can introduce more useful information for diagnosis. And the horizontal and vertical formulation combined with the vision transformer allows the model to analyze not only the tumor region itself but also the spatial relationship between the lesion and the different layers in the breast.

\begin{figure}[t]
	\includegraphics[width=.99\linewidth]{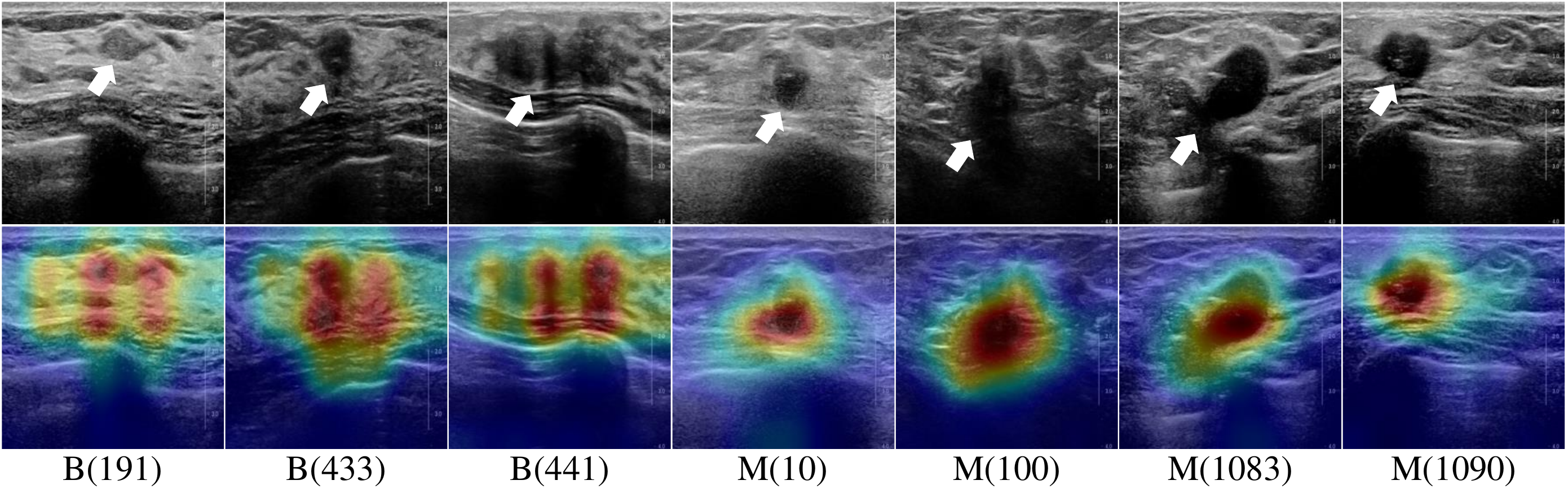}
	\caption{More visualization of HoVer-Trans. The top row shows the original BUS images in GDPH\&SYSUCC. The bottom row shows the heatmaps overlaid on the original images. M and B indicate the images with malignant tumors and benign tumors, respectively. The lesions are pointed by the white arrows.}
	\label{fig:featuremaps}
\end{figure}

Besides the quantitative results, we also demonstrate where the models focus by showing the heatmaps to evaluate the interpretability of the models (in the GDPH\&SYSUCC dataset). Fig.~\ref{fig:fm_compare} shows the heatmaps overlaid on the original images. We point out the lesions on the original image by white arrows in Fig.~\ref{fig:fm_compare}~(a) for better visualization. As can be seen, two transformer-based models in Fig.~\ref{fig:fm_compare}~(e)\&(f) cannot focus on the lesions and contains a lot of false-positive highlighted areas, which lead to poor specificity (TNT-s: 0.618, ViT: 0.694) shown in Table~\ref{tab:quantitative}. Visualization of two CNN-based models ResNet50 and VGG16 are shown in Fig.~\ref{fig:fm_compare}~(c)\&(d). ResNet50 also has the same problem with transformer-based models. VGG16 can achieve better visualization results compared with the previous three models. But it also pays attention to the dark areas caused by signal attenuation. Our proposed model in Fig.~\ref{fig:fm_compare}~(b) shows the best visualization results with more accurate lesion locations and more focused attention, thus achieving the best F1-score of 0.916. Fig.~\ref{fig:featuremaps} demonstrates more heatmaps of the HoVer-Trans model in GDPH\&SYSUCC. 

\vspace{-3mm}
\subsection{HoVer-Trans vs. Sonographers}\label{sec:AIvsHuman}
A reader study is conducted to compare the classification performance of HoVer-Trans with that of the experienced sonographers (YL and YW, at least ten years experience), one from the Department of Ultrasound, Guangdong Provincial People's Hospital and the other from the Department of Medical Ultrasonics, The First Affiliated Hospital of Guangzhou Medical University. The entire GDPH\&SYSUCC dataset is presented to the readers in a random order to assess the benignity or malignancy of the BUS lesion. To compare the entire dataset, we aggregate the model prediction results of each fold in the five-fold cross-validation.

\begin{table}[t]
	\centering
	\caption{HoVer-Trans vs. sonographers in the entire GDPH\&SYSUCC dataset and two subgroups. The last column shows the p-value of DeLong's test between the AUC of each reader and HoVer-Trans. p-values less than 0.001 are marked as ***.}
	\label{tab:quantitative_birads}
	\begin{tabular}{c|cccccc|c}
		\toprule[2pt]
		\multicolumn{8}{c}{GDPH\&SYSUCC}\\
		\hline
		& AUC & ACC & Spec & Prec & Rec & F1 & $p$-value\\
		\hline
		reader1     &0.825           &0.836           &0.781           &0.872           &0.868           &0.870&***\\
		reader2     &0.820           &0.838           &0.751           &0.859           &0.889           &0.874&***\\
		Ours        &\textbf{0.881} &\textbf{0.893} &\textbf{0.836} &\textbf{0.906} &\textbf{0.926} &\textbf{0.916}&-\\
		\toprule[2pt]
		\multicolumn{8}{c}{BI-RADS 2-3}\\
		
		\hline
		reader1     &0.503           &0.867           &\textbf{0.996}           &0.250           &0.099           &0.019&***\\
		reader2     &0.669           &0.828           &0.883           &0.368           &0.455           &0.407&***\\
		Ours        &\textbf{0.845}               &\textbf{0.870}	      &0.879	       &\textbf{0.5}	        &\textbf{0.812}	         &\textbf{0.619}&-\\
		\toprule[2pt]
		\multicolumn{8}{c}{BI-RADS 4-5}\\
		
		\hline
		reader1     &0.510           &0.821           &0.090           &0.873           &0.929           &0.901&***\\
		reader2     &0.622           &0.843           &0.324           &0.902           &0.920           &0.911&***\\
		Ours        &\textbf{0.817}               &\textbf{0.904}			  &\textbf{0.700}		   &\textbf{0.955}			&\textbf{0.934}			 &\textbf{0.944}&-\\
		\toprule[2pt]
	\end{tabular}
\end{table}

The uppermost part in Table~\ref{tab:quantitative_birads} shows the comparison between our model and two readers in all the evaluation metrics in the entire GDPH\&SYSUCC dataset. Experimental results show that our proposed model achieves more precise diagnostic performance than two readers in our dataset. However, we also observe that the diagnostic performance of two readers is lower than the one reported in the other reference~\cite{shen2021artificial}. Because in this experiment, to make a fair comparison with the AI model, readers assess only one image each time. But the ultrasound examination is a dynamic process, sonographers do not just read the static images but observe the lesion from different views. Therefore, it is hard for sonographers to precisely diagnose breast lesions with only one image. It is also a limitation of this experiment. Nevertheless, our proposed model outperforms sonographers under the same condition.

Furthermore, we conduct BI-RADS subgroup analysis in the lower part in Table~\ref{tab:quantitative_birads}. We divide the entire dataset into two subgroups by the BI-RADS scores, BI-RADS 2\&3 and BI-RADS 4\&5. It can be observed that the consistency between two readers is low reflected by the evaluation metrics, for example, the specificity (reader1: 0.996, reader2: 0.879) and the recall (reader1: 0.099, reader2: 0.455) in the BI-RADS 2\&3 group. Such biases can be caused by different factors, such as equipment bias, cognitive bias and etc. In addition, missing clinical information may also impede the sonographers to achieve a comprehensive diagnosis. Reader1 obviously tends to classify all the lesions with BI-RADS 2 or 3 as benign tumors, which may lead to undertreatment. Under the same condition of assessing only one BUS image without any additional information, our proposed model achieves more stable diagnostic performance in both subgroups than sonographers thanks to the data-driven nature.

\begin{table*}[htbp]
	\centering
	\caption{Ablation studies of anatomy-aware formulation, association between transformer and CNN and different transformer configurations. (GDPH\&SYSUCC)}
	\label{tab:ablation}
	\begin{tabular}{cc|ccccccc}
		\toprule[2pt]
		\multicolumn{8}{c}{Ablation study - Anatomy-aware formulation}\\
		\hline
		\multicolumn{2}{c|}{Configurations} & AUC & ACC & Specificity & Precision & Recall & F1-score\\
		\hline
		\multicolumn{2}{c|}{$Model_p$}     &0.916$\pm$0.018 &0.864$\pm$0.015 &0.802$\pm$0.016 &0.887$\pm$0.012 &0.900$\pm$0.016 &0.893$\pm$0.013\\
		\multicolumn{2}{c|}{$Model_{p+v}$} &0.919$\pm$0.022 &0.880$\pm$0.018 &\textbf{0.849$\pm$0.037} &\textbf{0.910$\pm$0.023} &0.898$\pm$0.018 &0.904$\pm$0.015\\
		\multicolumn{2}{c|}{$Model_{p+h}$} &0.911$\pm$0.019 &0.868$\pm$0.019 &0.818$\pm$0.021 &0.894$\pm$0.014 &0.897$\pm$0.022 &0.896$\pm$0.015\\
		\multicolumn{2}{c|}{Ours}  &\textbf{0.924$\pm$0.016} &\textbf{0.893$\pm$0.021} &0.836$\pm$0.038 &0.906$\pm$0.023 &\textbf{0.926$\pm$0.019} &\textbf{0.916$\pm$0.019}\\
		
		\toprule[2pt]
		\multicolumn{8}{c}{Ablation study - Conv block}\\
		\hline
		\multicolumn{2}{c|}{Configurations} & AUC & ACC & Specificity & Precision & Recall & F1-score\\
		\hline
		\multicolumn{2}{c|}{w/o Conv}  &0.916$\pm$0.016 &0.873$\pm$0.012 &\textbf{0.900$\pm$0.011} &0.827$\pm$0.026 &0.826$\pm$0.013 &0.826$\pm$0.010\\
		\multicolumn{2}{c|}{with Conv $1\times 1$}	&\textbf{0.926$\pm$0.015} &0.881$\pm$0.018 &0.837$\pm$0.031 &0.905$\pm$0.018 &0.907$\pm$0.017 &0.905$\pm$0.015\\
		\multicolumn{2}{c|}{with Conv (Ours)} &0.924$\pm$0.016 &\textbf{0.893$\pm$0.021} &0.836$\pm$0.038 &\textbf{0.906$\pm$0.023} &\textbf{0.926$\pm$0.019} &\textbf{0.916$\pm$0.019}\\	
		
		\toprule[2pt]
		\multicolumn{8}{c}{Ablation study - Sizes of Different Embedding Ways}\\
		\hline
		$p$ & $h\& v$ & AUC & ACC & Specificity & Precision & Recall & F1-score\\
		\hline
		2 & 1   &0.911$\pm$0.024 &0.880$\pm$0.014 &0.826$\pm$0.025 &0.900$\pm$0.014 &0.912$\pm$0.014 &0.906$\pm$0.011\\
		2 & 2   &\textbf{0.924$\pm$0.016} &\textbf{0.893$\pm$0.021} &0.836$\pm$0.038 &0.906$\pm$0.023 &\textbf{0.926$\pm$0.019} &\textbf{0.916$\pm$0.019}\\
		2 & 4   &0.910$\pm$0.016 &0.873$\pm$0.016 &0.825$\pm$0.029 &0.899$\pm$0.014 &0.900$\pm$0.020 &0.900$\pm$0.012\\					
		4 & 1   &0.904$\pm$0.027 &0.877$\pm$0.021 &0.837$\pm$0.026 &0.905$\pm$0.015 &0.900$\pm$0.021 &0.903$\pm$0.016\\
		4 & 2   &0.919$\pm$0.015 &0.883$\pm$0.014 &\textbf{0.841$\pm$0.017} &\textbf{0.908$\pm$0.010} &0.907$\pm$0.013 &0.907$\pm$0.011\\
		4 & 4   &0.920$\pm$0.014 &0.879$\pm$0.012 &0.832$\pm$0.030 &0.903$\pm$0.017 &0.906$\pm$0.014 &0.905$\pm$0.010\\
		8 & 1   &0.911$\pm$0.013 &0.879$\pm$0.010 &0.835$\pm$0.020 &0.904$\pm$0.014 &0.905$\pm$0.008 &0.904$\pm$0.009\\
		8 & 2   &0.914$\pm$0.019 &0.879$\pm$0.014 &0.820$\pm$0.023 &0.897$\pm$0.016 &0.913$\pm$0.011 &0.905$\pm$0.012\\
		8 & 4   &0.900$\pm$0.011 &0.864$\pm$0.008 &0.792$\pm$0.019 &0.882$\pm$0.010 &0.906$\pm$0.015 &0.894$\pm$0.007\\
		\toprule[2pt]
	\end{tabular}
\end{table*}

\subsection{Ablation Studies}\label{sec:exp_ablation}
In this part, we conduct several ablation studies to evaluate the effectiveness of the anatomy-aware formulation, the association between transformer and CNN and different transformer configurations.

\subsubsection{Effectiveness of Anatomical Prior Knowledge}
We conduct this experiment to evaluate the effectiveness of the anatomy-aware formulation. Three variants are introduced in this experiment. 1) H branch with horizontal strip embedding and V branch with vertical strip embedding are removed, named $Model_p$. Only two main branches with patch embedding are left. 2) We remove the H branch and retain the other three branches, named $Model_{p+v}$. 3) We remove the V branch and retain the other three branches, named $Model_{p+h}$. 
The upper part of Table~\ref{tab:ablation} demonstrates the five-fold cross validation results. 
It can be observed that without the design of HoVer in $Model_p$, the performance of the metrics decreases by around 1\%-3\% except for the specificity and precision of $Model_{p+v}$. When only removing H branch or V branch, the quantitative results do not improve due to the asymmetric of the models. Fig.~\ref{fig:ablation}~(b)-(d) demonstrate the heatmap visualizations of three model variants and Fig.~\ref{fig:ablation}~(g) demonstrates our results. In Fig.~\ref{fig:ablation}~(b) we can observe that associating transformer with CNN can obtain visually more convincing attention maps better than transformer models only, shown in Fig.~\ref{fig:featuremaps}~(e)\&(f). However, the model focuses are still imperfect due to the lack of H and V branches. When equipped with H or V branch in Fig.~\ref{fig:ablation}~(c)\&(d), the model can focus on the anomaly regions horizontally or vertically, guided by the anatomical prior. 
Thanks to the complete HoVer design shown in Fig.~\ref{fig:ablation}~(g), our proposed model achieves the best visualization results with the most correct lesion location and attention.

\subsubsection{Effectiveness of Conv Block}
\begin{figure*}[t]
	\includegraphics[width=.99\linewidth]{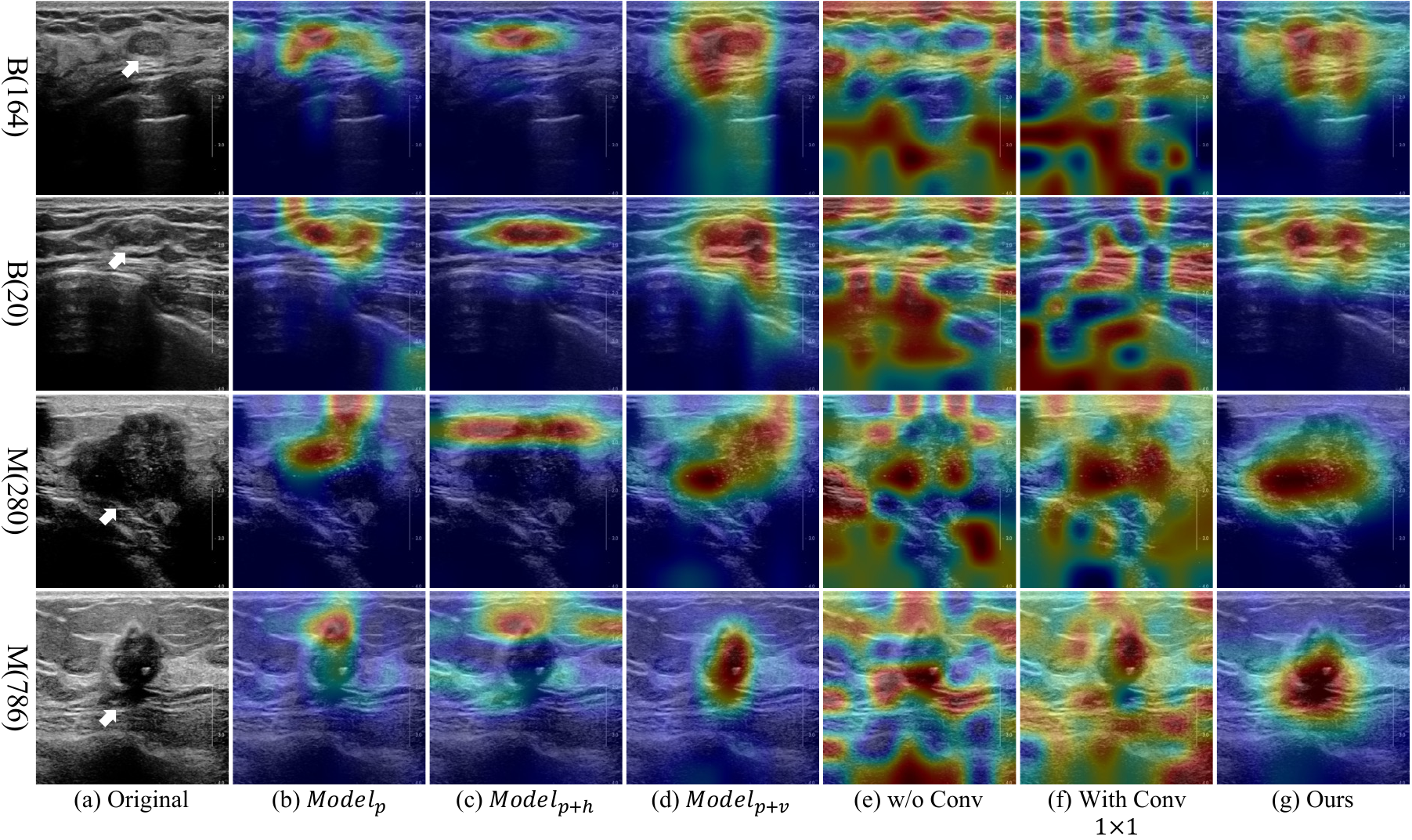}
	\caption{The heatmaps of different convolutional layers (ablation study). (a) shows the original images from the GDPH\&SYSUCC dataset. (b)-(d) are the model variants of the ablation study in the anatomy-aware formulation. (e) is the model with pooling only without Conv block. (f) is the model with $1\times 1$ convolutional layer. (e) is our final model with Conv block and HoVer design.}
	\label{fig:ablation}
\end{figure*}

In this study, we utilize the convolutional block to connect two adjacent HoVer-Trans stages to introduce multi-scale model representation and inductive bias. In order to evaluate the advantage of associating transformer with CNN, we compare our model with two variants. 1) We remove the Conv block and replace it with an average pooling layer for feature dimension reduction (w/o Conv). 2) We replace the Conv block with a $1\times 1$ convolutional layer. The quantitative results are shown in the middle part in Table~\ref{tab:ablation}. The attention maps are shown in Fig.~\ref{fig:ablation}~(e)-(g).

When without the convolutional layer, we can observe a decreased quantitative performance and the attention maps totally lost focus. Introducing a $1\times 1$ convolutional layer can slightly improve the classification performance, but the attention maps are still unsatisfactory (Fig.~\ref{fig:ablation}~(f)). With the interaction of horizontal and vertical anatomy-aware formulation, the proposed HoVer-Trans model can achieve the best classification results and the most interpretable attention maps.

\subsubsection{Sizes of Different Embedding Ways}
Since we introduce three embedding ways in this model to formulate the anatomical structure. In this ablation study, we further explore how the sizes of different embedding ways affect the proposed model, shown in Table~\ref{tab:ablation}. $p=2$ means the visual tokens of the patching embedding are with the resolution of $2\times 2$. $h\& v=2$ means the tokens of the horizontal strip embedding and the vertical strip embedding are with the resolution of $2\times width$ and $height\times 2$, respectively. In this experiment, we let $p=\{2,4,8\}$ and $h\&v = \{1,2,4\}$ and test all the combinations. 

It can be observed that the classification performance of all the models is close. According to the quantitative results in Table~\ref{tab:ablation}, we summarize some observations on how to select these two hyper-parameters. First of all, we have to select the proper token size of the horizontal and vertical strip embeddings $h\&v$. When $h\&v=1$, each token only contains the horizontal or vertical information with only one-pixel width or height, which is too limited. When $h\&v=4$, the size of the token is too large, which may occupy more computational resources. Therefore, we let $h\&v=2$ for the horizontal and vertical strip embedding. For the token size of patch embedding, we choose the value of $p=2$ because we want to let the token of patch embedding fit the token size of horizontal and vertical strip embedding. Experimental results prove that the configuration with $p=2$ and $h\&v=2$ achieves the best classification performance. We apply this setting in our final model.

\section{Conclusion}\label{sec6}
In this paper, we propose a novel HoVer-Trans model, which associates the transformer with CNN, for breast cancer diagnosis in breast ultrasound images. An anatomy-aware HoVer-Trans block is designed to formulate the anatomical prior knowledge in BUS images. To achieve that, we incorporate three embedding ways, patch embedding, horizontal strip embedding and vertical strip embedding to explore spatial correlations of the inter-layer and intra-layer visual words. There are several advantages to the above technical designs. 1) The proposed model is ROI-free which does not require a pre-defined lesion ROI. Such a property greatly improves the model flexibility in clinical practice. We also believe that the whole image can deliver much more information about the peritumoral context and the spatial relationship between the lesion and different breast layers than the lesion ROIs do. 2) The proposed model can provide interpretable attention maps to support the model prediction results, which is the key that most sonographers care about when using AI algorithms to assist decision-making. 3) The proposed model achieves the best classification performance against several SOTA models in both quantitative evaluations and heatmap visualizations.

Besides, there are still a lot of future improvements that have to be achieved.
First, as we discussed in the AI vs. sonographers experiment, the breast ultrasound examination is a dynamic process. Our next plan is to aggregate the BUS images from multiple views and achieve more precise diagnostic performance, instead of just simply assessing one BUS image. Furthermore, it will be a great breakthrough if we can mimic how sonographers perform breast ultrasound examination by keeping tracking the imaging signal along with the ultrasonic probe and make a comprehensive diagnosis if the hardware is capable.

Second, due to the model complexity, the proposed model shows poor classification performance when trained by a smaller dataset, such as UDIAT. That is the reason why we also construct and release a larger dataset GDPH\&SYSUCC for breast cancer diagnosis in BUS images. We are also planning to construct a way larger multi-center dataset to further explore the capacity and the generalizability of the proposed model.

%
\bibliography{ref}
\bibliographystyle{IEEEtran}

\end{document}